\documentclass{mn2e}

\usepackage{aas_macros}
\usepackage{graphicx}
\usepackage{amsmath}
\usepackage{natbib}
\usepackage{xspace}

\usepackage{array}
\newcolumntype{L}[1]{>{\raggedright\let\newline\\\arraybackslash\hspace{0pt}}p{#1}}
\newcolumntype{C}[1]{>{\centering\let\newline\\\arraybackslash\hspace{0pt}}p{#1}}
\newcolumntype{R}[1]{>{\raggedleft\let\newline\\\arraybackslash\hspace{0pt}}p{#1}}

{ 

\newcommand{\chisq}{$\chi ^2$\xspace}

\usepackage[utf8]{inputenc}
\usepackage{color}

\title[Secondary eclipses of WASP-18b]{Secondary eclipses of WASP-18b -- Near Infrared observations with the Anglo Australian Telescope, the Magellan Clay Telescope and the LCOGT network}

\author[L.~Kedziora-Chudczer et al.]
{\parbox{\textwidth}
{L.~Kedziora-Chudczer$^1$, 
G.~Zhou$^2$, J.~Bailey$^1$, D.D.R.~Bayliss$^3$, C.G.~Tinney$^4$, D.~Osip$^{5}$, K.D.~Col\'{o}n$^{6}$, A.~Shporer$^{7}$, D.~Dragomir$^{7}$\thanks{NASA Hubble Fellow}
\vspace{0.4cm}}\\
\parbox{\textwidth}{
$^{1}${School of Physics, UNSW Sydney, NSW, 2052, Australia}\\
$^{2}${Harvard-Smithsonian Center for Astrophysics, 60 Garden St, Cambridge MA, USA, 02138 }\\
$^{3}${Department of Physics, University of Warwick, Coventry CV4 7AL, UK}\\
$^{4}${Exoplanetary Science at UNSW, School of Physics, UNSW Sydney, NSW, 2052, Australia}\\
$^{5}${Las Campanas Observatory Carnegie Institution of Washington
Colina El Pino Casilla 601 La Serena, Chile}\\
$^{6}${NASA Goddard Space Flight Center, Exoplanets and Stellar Astrophysics Laboratory (Code 667), Greenbelt, MD 20771, USA}\\
$^{7}${Massachusetts Institute of Technology, Cambridge, MA 02139, USA}}}

\begin{document}

\date{Accepted 7/12/2018.  Received XXX; in original form XXX}

\pagerange{\pageref{firstpage}--\pageref{lastpage}} \pubyear{2015}

\maketitle

\label{firstpage}

\begin{abstract}
We present new eclipse observations for one of the hottest ``hot Jupiters'' WASP-18b, for which previously published data from HST WFC3 and Spitzer have led to radically conflicting conclusions about the composition of this planet's atmosphere.  We measure eclipse depths of $0.15\pm0.02\%$ at $Ks$ and $0.07\pm0.01\%$ at $z'$ bands.
Using the VSTAR line-by-line radiative transfer code and both these new observations with previously published data, we derive a new model of the planetary atmosphere.
We have varied both the metallicity and C/O ratio in our modelling, and find no need for the extreme metallicity suggested by \citet{2017ApJ...850L..32S}. 
\color{black} Our best fitting models slightly underestimate the emission at $z'$ band and overestimate the observed flux at $Ks$-band. To explain these discrepancies, we examine the impact on the planetary emission spectrum of the presence of several types of hazes which could form on the night-side of the planet.  Our $Ks$ band eclipse flux measurement is lower than expected from clear atmosphere models and this could be explained by a haze particles larger than 0.2\,$\mu$m with the optical properties of Al$_{2}$O$_{3}$, CaTiO$_{3}$ or MgSiO$_{3}$. We find that $z'$ band measurements are important for understanding the contribution of photochemical hazes with particles smaller than 0.1\,$\mu$m at the top of the atmosphere.\color{black} 
\end{abstract}


\section{Introduction}
\label{sec:introduction}
WASP-18b is a massive, hot Jupiter in a very short-period orbit around an F6V star \citep{2009Natur.460.1098H}. 
Although a handful of planetary systems with orbital periods of under a day are known, WASP-18b is the only high-mass example (i.e. more than 10 Jupiter masses) of such a system. This makes it a good candidate for study of potentially expected strong tidal interactions with the parent star \citep{2017ApJ...836L..24W}. Such interactions should lead to mass loss and orbital instability, resulting in a relatively short life span of the planet. Indeed, the shortening of its orbital period may be detectable within just a few years. 

WASP-18b's very strong stellar irradiation is also expected to influence its atmospheric structure leading to an inverted thermal profile with high stratospheric temperatures \citep{2008ApJ...678.1419F} -- a mechanism that may contribute to the observed inflated radii of many hot Jupiters \citep{2010RPPh...73a6901B}. Models of the planetary spectrum for WASP-18b \citep{2011ApJ...742...35N} show a slightly better fit to the planet's day-side spectrum for atmosphere profiles with temperature inversion as a function of pressure, compared to models without such an inversion. 


WASP-18b was discovered as a primary transiting system, and its orbital properties have been characterised by \citet{2009ApJ...707..167S} from monitoring observations in the V band at the 1.5m Danish Telescope at ESO's La Silla Observatory. The first observations of the WASP-18b's secondary eclipses were obtained with the IRAC camera on the {\em Spitzer} telescope \citep{2011ApJ...742...35N}., with brightness temperatures in the four spectral channels of IRAC (3.6, 4.5, 5.8, 8.0\,$\mu$m)  above 3000\,K making WASP-18b one of the most irradiated hot Jupiters. 

New HST observations with the Wide Field Camera (WFC3) have been published independently by both
\citet{2017ApJ...850L..32S} and \citet{2018ApJ...855L..30A}. The models of atmospheric composition derived by these studies, however, are radically different, requiring  either extremely high metallicities of $C/H=283$S$_{\odot}$ and $C/O=1$  \citep{2017ApJ...850L..32S}, or Solar values for both parameters \citep{2018ApJ...855L..30A}. \color{black} It is worth noting the latter group considered a more sophisticated model that included opacity sources due to thermal ionisation and dissociation of atmospheric species that are important at high temperatures \citep{2018A&A...617A.110P}. \color{black}


The modelling of exoplanetary atmospheres remains challenging. Currently it is impossible to acquire simultaneous data across a wide spectral range, and this, in turn, makes it impossible to resolve many model degeneracies. For example, a time-dependent radiative transfer model \citep{2013Icar..226.1719I} proposed to examine heat transfer efficiency in WASP-18b shows differences in spectra observed at different phases when compared with static averaged models. The planetary rotation period, often difficult to estimate, will influence the speed of equatorial winds and these transfer heat from the sub-stellar (i.e. star-facing) region. This will tend to level out the planetary fluxes measured in different orbital phases.  

In addition most \color{black} atmospheric retrieval models \color{black} for hot Jupiters is done using simplified chemical assumptions that ignore photochemical processes (like the photodissociation of water), which may change the expected levels of molecular absorption of H$_{2}$O in observed spectra \citep{2012ApJ...745...77K}. Further complexity in modelling of such atmospheres is the suggested presence of hazes or clouds that appear to fit the observed transit data \citep[the best known example of which is HD 189733b --][]{2014ApJ...791...55M,2013MNRAS.432.2917P}. The types of hazes that could form at the top layers of a hot Jupiter's atmosphere are discussed by \citet{2013cctp.book..367M} and \citet{2016Natur.529...59S}. Models that include hazes in the radiative transfer solutions, require the availability of  optical property data for the refractory particles that could survive in such hot atmospheres. For the hottest hot Jupiters (like WASP-18b) the condensation temperature of the most refractory chemicals are lower than the blackbody temperature of the planet. \color{black} Therefore it is not clear if {\em any} clouds or hazes can be formed and long-lived in the planetary atmosphere of WASP-18b. Modelling by \citet{2016ApJ...828...22P} suggest that a fraction of the planet's day side, along the western terminator could remain covered by condensate clouds even in high equilibrium temperatures. Similarly the results of the 3-D dynamical models for other highly irradiated, hot Jupiter, WASP-121b \citep{2017ApJ...845L..20K} suggest that any condensation that would occur on the dark side of the planet would be destroyed relatively quickly during circulation toward the sub-stellar regions of the atmosphere, however the photochemical hazes formed on the day side could be transported and survive in all regions of the planet. \color{black}

Here we examine models with (potentially transient) hazes of different composition and opacity. The models are constrained by our new observations of WASP-18b eclipses at $K_{s}$ and $z'$ bands, which attempt to fill in the gaps in previously observed spectral regions  (Section \ref{sec:observations}). We improve these constraints for the modelled spectra by combining our data with previously published observations.   \color{black} Accurate $K_{s}$ and $z'$ band depths extend the coverage of emission spectrum on both sides of the range accessible to the WFC3 camera. Both measurements can provide a signature of the presence of condensate clouds and photochemical hazes in the atmosphere, as discussed later.
 Furthermore, observations in these two bands can be particularly useful in the statistical studies on hot Jupiters, when the distribution of $K_{s}$ and $z'$ band observations for similar planets, becomes a useful guide to promote in-depth characterisation by space-based facilities. \color{black}

 In Section~\ref{sec1} we derive the best fitting model of the emission spectrum with the Versatile Software for Transfer of Atmospheric Radiation (VSTAR) \citep{2012MNRAS.419.1913B}. Our atmosphere model includes many opacity sources omitted in \citet{2017ApJ...850L..32S}. Finally we focus our discussion on the different types of hazes that could be present, at least intermittently, in the atmosphere of WASP-18b. We model the emission spectrum with the addition of hazes with different optical depths and particle sizes to examine features in different parts of the mid- and far-infrared spectrum (Section\ref{sec:haze}) that will be soon accessible to observations with the James Webb Space Telescope (JWST).

\section{Observations and analysis}
\label{sec:observations}


We obtained a series of secondary eclipse observations of WASP-18b at multiple near-infrared wavelengths, including 4 epochs of $z'$ band eclipses from the Las Cumbres Observatory Global Telescope (LCOGT) 1\,m network, 1 epoch of $z'$ band data from the 6.5\,m Magellan Clay telescope at Las Campanas Observatory (LCO), and 2 epochs of $Ks$ band eclipses from the 3.9\,m Anglo-Australian Telescope (AAT). We also observed 2 epochs of $r'$ band primary transits with LCOGT used in the analysis discussed below. The observations are summarised in Table~\ref{tab:observations_list}.

\begin{table*}
  \centering
  \caption{Table of the observations}
  \label{tab:observations_list}
  \begin{tabular}{lllrrrrr}
    \hline\hline
    Facility & Date & Type & band & Number & Cadence & Exposure & Median \\
    & (UT) & & & of Exps & (s) & Time (s) & FWHM (pix)\\
    \hline
    AAT+IRIS2 & 2014-09-05 & Secondary & $Ks$ & 95 & 120$^a$ & 5 & 19 \\
    AAT+IRIS2 & 2015-09-27 & Secondary & $Ks$ & 3832 & 2.6 & 2 & 7 \\
    LCOGT 1\,m+Sinistro (CTIO) & 2015-11-02 & Secondary & $z'$ & 201 & 79 & 40 & 38 \\
    LCOGT 1\,m+Sinistro (CTIO) & 2015-11-04 & Secondary & $z'$ & 163 & 79 & 40 & 27 \\
    LCOGT 1\,m+Sinistro (CTIO) & 2015-11-11 & Primary & $r'$ & 185 & 49 & 10 & 37 \\
    LCOGT 1\,m+Sinistro (CTIO) & 2015-11-12 & Primary & $r'$ & 260 & 49 & 10 & 53 \\
    LCOGT 1\,m+Sinistro (CTIO) & 2015-11-19 & Secondary & $z'$ & 202 & 79 & 40 & 37 \\
    LCOGT 1\,m+Sinistro (CTIO) & 2015-11-20 & Secondary & $z'$ & 158 & 79 & 40 & 53 \\
    Magellan+LDSS3 & 2016-09-17 & Secondary & $z'$ & 431 & 45 & 10 & 220 \\
    \hline
  \end{tabular}
\begin{flushleft} 
$20 \times 5s$ exposures are averaged and used for analysis. Individual exposures were not saved. Observation presented in \citet{2015MNRAS.454.3002Z}.
\end{flushleft}
\end{table*}

\begin{figure*}
    \centering
    \includegraphics[width=15cm]{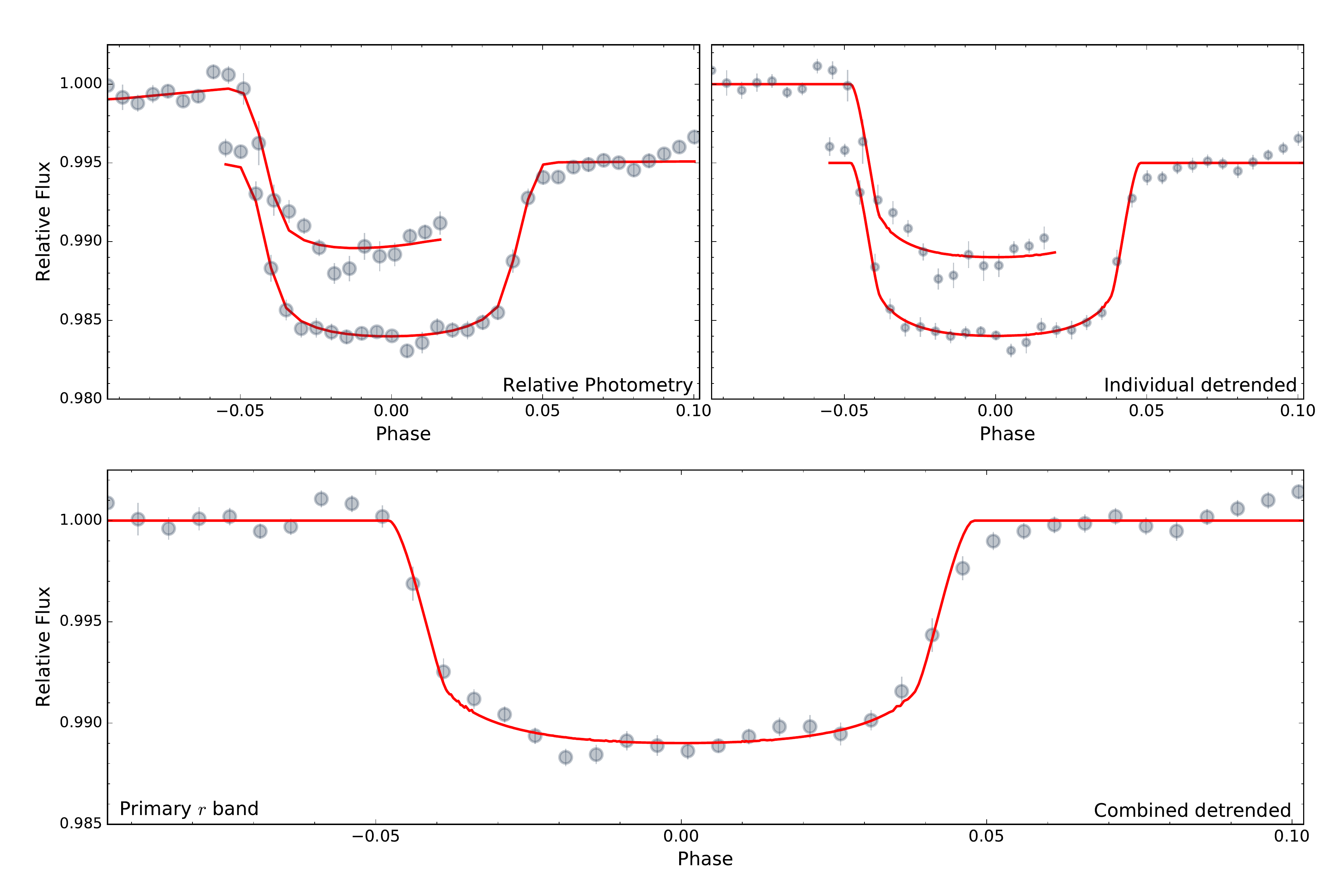}
    \caption{The LCOGT 1\,m $r$ band primary transits of WASP-18b. The top left panel shows light curves before de-trending for each observation, binned at 0.005 in phase. The uncertainties indicate the standard deviation of the points within each bin, scaled by the square root of the number of points in the bin. Successive light curves are offset arbitrarily by 0.005 in flux for clarity. The red line shows the best fit model (transit and instrumental). The top right panel shows each light curve after simultaneous de-trending after the global fit. The best fit transit model is plotted in red. The bottom panel shows the combined de-trended light curve.}
    \label{fig:primary_joint}
\end{figure*}

Primary transits were observed on 2015-11-11 (partial) and 2015-11-12 (full) with the LCOGT 1\,m network, from Cerro Tololo Inter-American Observatory (CTIO), Chile. The observations were obtained with the $4\text{K}\times4\text{K}$ Sinistro cameras, using a Fairchild CCD-486 back-illuminated detector with a field of view of $26'\times26'$ at $0.389''/\text{pixel}$. The observations were performed in the $r'$ photometric band with exposure times of 10\,s. In each case, the telescope was highly defocused, yielding point-spread-functions with full width at half maxima of 30-40 pixels. The primary transit light curves are shown in Figure~\ref{fig:primary_joint}. 

Four full secondary eclipses WASP-18b were also observed with the 1\,m LCOGT telescope at CTIO in the $z'$ band with the Sinistro cameras, on 2015-11-02, 2015-11-04, 2015-11-19 and 2015-11-20. As for the primary transit observations, the eclipse sequences were performed with the telescope defocused, but with longer exposure times of 40\,s. 

Bias subtraction and flat fielding were performed automatically by the LCOGT pipeline, and photometry of the target star and selected reference stars were extracted from these frames. First, we solve for the astrometric solution of each frame to derive the centroids of extraction. Aperture photometry is then performed using the \emph{fitsh} package \citep{2012MNRAS.421.1825P} over six fixed aperture sizes. The background flux is estimated by the median flux within an annulus surrounding the photometric aperture. 

\begin{figure*}
    \centering
    \includegraphics[width=15cm]{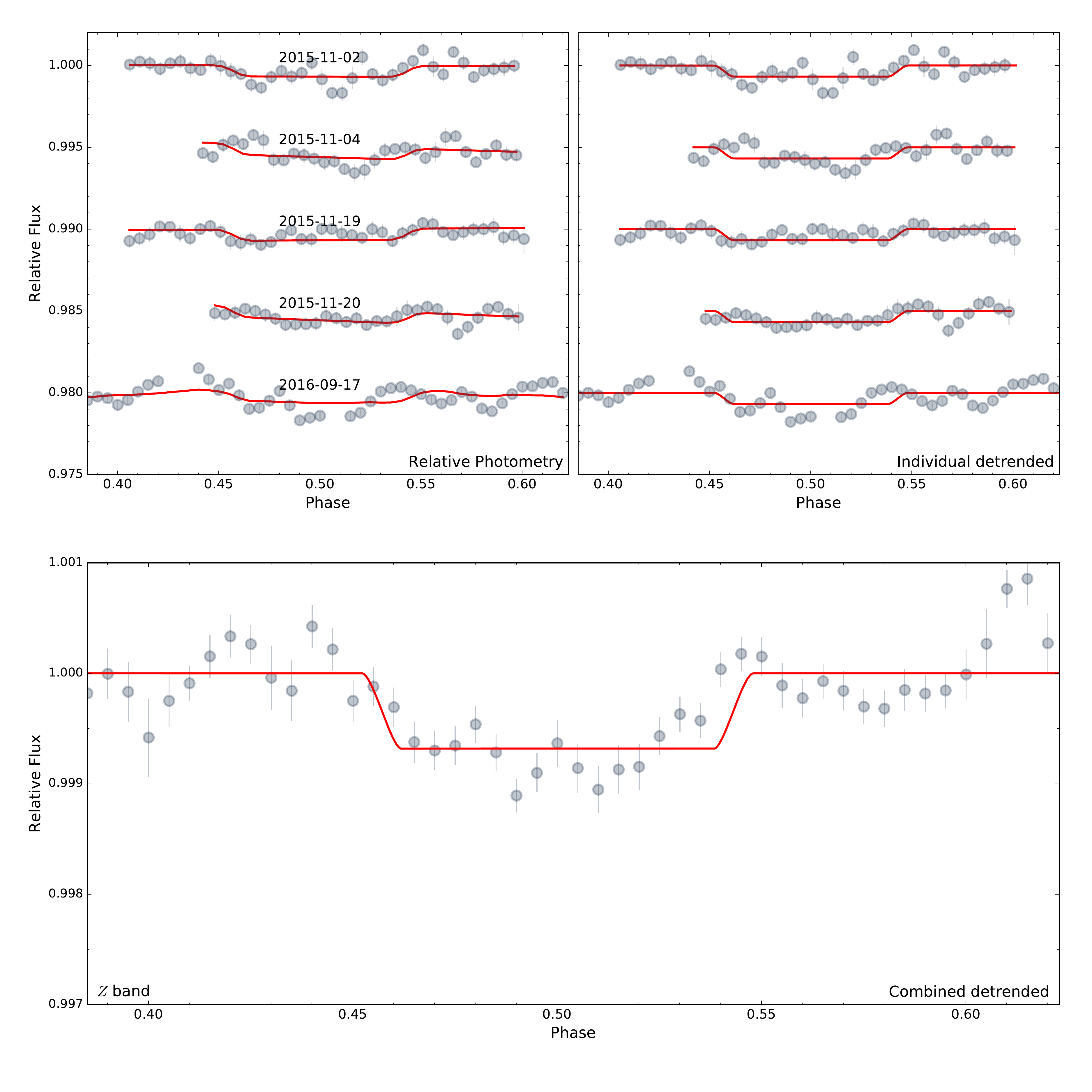}
    \caption{Four LCOGT 1\,m $z'$ band secondary eclipses were obtained between 2015-11-02 and 2015-11-20. An additional $z'$ band full eclipse was obtained with Magellan LDSS3 on 2016-09-17. The plot formats are as per Figure~\ref{fig:primary_joint}.}
    \label{fig:Z_joint}
\end{figure*}

One additional $z'$ band secondary eclipse was observed with the 6.5\,m Magellan Clay telescope at LCO on 2016-09-17 with the Low Dispersion Survey Spectrograph 3 (LDSS3) instrument. LDSS3 uses a  S3-171 $2\,K\times4\,K$ red-sensitive detector, with a circular field-of-view of $6'.4$ in diameter and a pixel scale of $0.189''\,\mathrm{pixel}^{-1}$. A total of 431 observations were obtained with a 10\,s exposure time and the target severely defocused to a full width half maximum of $\sim 200$ pixels. The observations were bias subtracted, and then flat fielded with twilight sky frames. Photometry from target and reference stars are extracted using \emph{fitsh}, as for the LCOGT observations. The $z'$ band secondary eclipses from LCOGT and LSDSS3 are shown in Figure~\ref{fig:Z_joint}.

Near-infrared secondary eclipses were observed with the IRIS2 camera on the 3.9\,m AAT, at Siding Spring Observatory, Australia. IRIS2 is a $1\text{K}\times1\text{K}$ infrared camera employing a HAWAII-1 HgCdTe infrared detector with four readout quadrants, and records in double-read mode. The IRIS2 installed on the AAT yields a $7'.7\times7'.7$ field of view and a plate scale of $0.4488''\,\mathrm{pixel}^{-1}$. 

\begin{figure*}
    \centering
    \includegraphics[width=15cm]{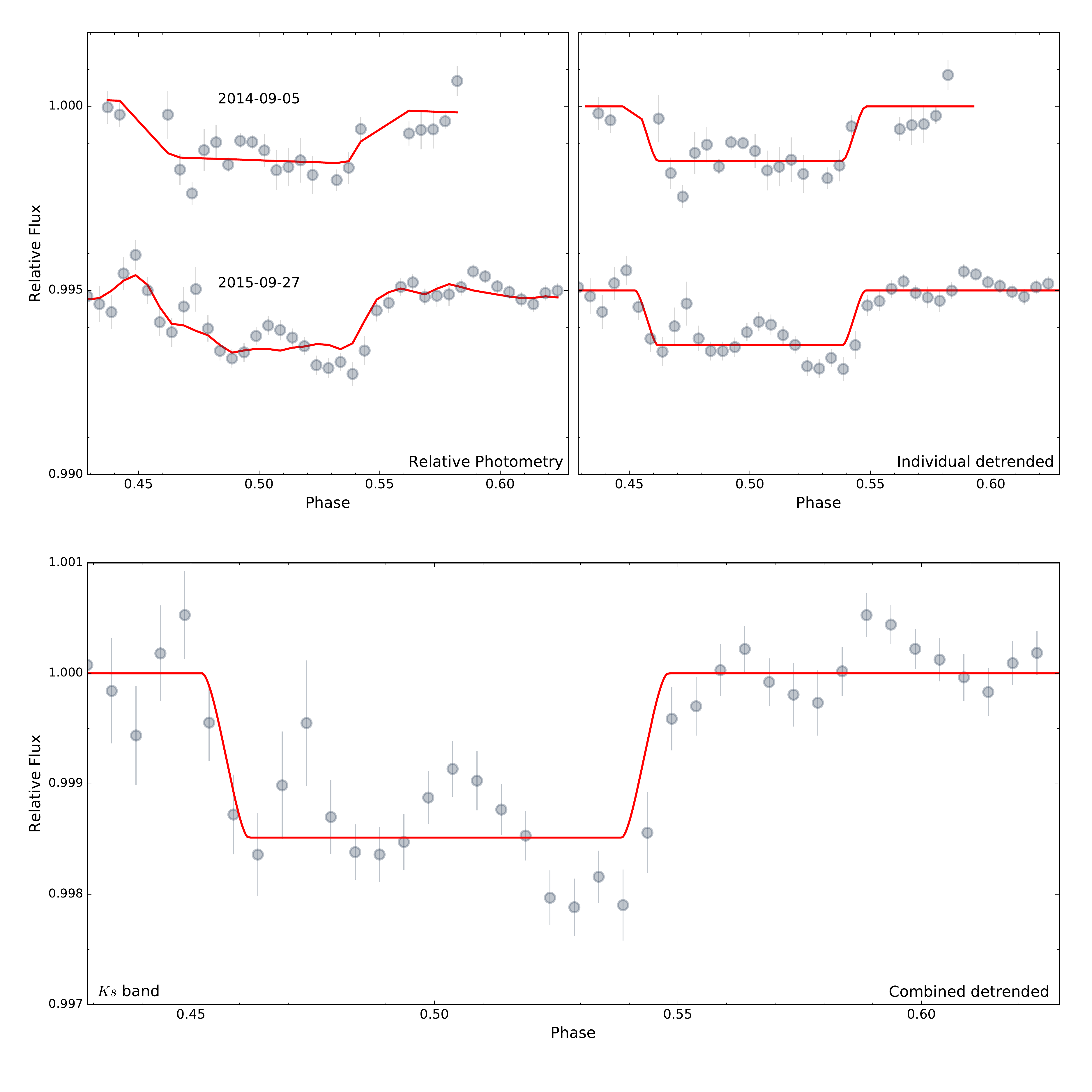}
    \caption{Two $Ks$ band secondary eclipses of WASP-18b were obtained by AAT+IRIS2 on 2014-09-05 and 2015-09-27. The plot formats are as per Figure~\ref{fig:primary_joint}.}
    \label{fig:Ks_joint}
\end{figure*}

The observing strategy and data reduction for the IRIS2 observations are laid out in \citet{2014MNRAS.445.2746Z} and \citet{2015MNRAS.454.3002Z}. Each eclipse sequence is observed in guided ``stare'' mode. The observer checks for tracking drifts in the science frames every $\sim 10$ minutes by solving the astrometry, and inputs manual tracking adjustments as necessary. As a result, the centroid of the target star remains on the same pixel throughout each observation sequence. Dark subtraction is performed on each science exposure. A master sky frame is assembled from sets of offset exposures that bracket each exposure sequence, and this is used to provide a flat-field for the science frames. Aperture photometry for the target star and selected reference stars is then performed on the reduced frames following the same procedure as for LCOGT observations. The IRIS2 $Ks$ band secondary eclipses are shown in Figure~\ref{fig:Ks_joint}.

\section{Light curve analysis and modelling}
\label{sec:light-curve-analysis}

\subsection{Global fit}
\label{sec:global-fit}

We perform a global fit of all our observations, including the LCOGT primary transits, as well as LCOGT and Magellan $z'$ band and AAT+IRIS2 $Ks$ band secondary eclipses data. The light curve modelling process largely follows that described in \citet{2014MNRAS.445.2746Z} and \citet{2015MNRAS.454.3002Z}. The transits and eclipse are modelled with the \emph{JKTEBOP} \citep{1981AJ.....86..102P,2004MNRAS.351.1277S} implementation of the \emph{EBOP} model \citep{1972ApJ...174..617N}. The free parameters governing the transit geometry are the planet-star radius ratio $R_p/R_\star$, normalised orbital distance $(R_p+R_\star)/a$, inclination $i$, orbital period $P$ and transit centre timing $T_0$. The parameters governing the secondary eclipse are the surface brightness ratios $S_{p}/S_\star$ and the eccentricity parameter $e\cos\omega$ (which dictates the phase of the eclipse).

In addition, we allow for simultaneous de-trending of the light curve against a combination of the external parameters: target star pixel position ($X$, $Y$) , airmass $a$, background flux $b$, target star FWHM $f$, and a linear time trend $t$. The combination of external parameters for each data set is determined from independent fits to each individual data set that minimise the Bayesian Information Criterion (BIC) of the model (see Section~\ref{sec:individual-fit}). This allows us to model systematic trends in the light curve due to atmospheric and instrumental variations. \color{black} To account for underestimated per-point photometric uncertainties in the light curves, we also include a separate photometric jitter term to each light curve, such that the log likelihood $\ln \mathcal{L}$ is calculate as,
\begin{equation}
    \ln \mathcal{L} = - \frac{n}{2} \ln(2\pi) - \frac{1}{2} \chi^2 - \frac{1}{2} \sum^{i=n}_{i=0} \ln (\sigma_\mathrm{lc}^2 + s_\mathrm{lc}^2)\,,
\end{equation}
where $n$ is the number of points per light curve, $\sigma_\mathrm{lc}$ is the photometric per point uncertainty, and $s_\mathrm{lc}$ is the photometric jitter term. \color{black}

These parameters and associated uncertainties are explored via an Markov-chain Monte Carlo analysis, using the \emph{emcee} Affine  Invariant Ensemble sampler \citep{2013PASP..125..306F}. Tight Gaussian priors are placed on the transit parameters $R_p/R_\star$, $(R_p+R_\star)/a$, $i$, $P$, and $T_0$ from literature system values \citep{2009ApJ...707..167S}. The derived parameters are presented in Table~\ref{tab:global_fit}.

\begin{table}
    \caption{\color{black}Global fitting parameters including photometric jitter terms\color{black}}
    \label{tab:global_fit}
    \centering
    \begin{tabular}{ll}
        \hline\hline
        Parameter & Value  \\
        \hline
        \emph{MCMC jump parameters} \\
        Period (d) & $0.94145180\,\left(45\right)$ \\
        $T_0$ (BJD-TDB) & $2455084.7948 \pm 0.0011$ \\
        $(R_p+R_\star)/a$ & $0.3049 \pm 0.0058$ \\
        $R_p/R_\star$ & $0.09868 \pm 0.00093$ \\
        $i$ $(^\circ)$ & $85.6 \pm 1.1$ \\
        $e\cos \omega$ & $0.0002 \pm 0.0016$ \\
        $S_p/S_\star$ $z'$-band & $0.070 \pm 0.010$\\
        $S_p/S_\star$ $Ks$-band & $0.153 \pm 0.015$\\
        $s_{AAT\,Ks\,20140905}$ & $0.00108\pm0.00011$\\
        $s_{AAT\,Ks\,20150927}$ & $0.00409\pm0.00051$\\
        $s_{LCO\,z'\,20151102}$ & $0.00145\pm0.00082$\\
        $s_{LCO\,z'\,20151104}$ & $0.00140\pm0.00093$\\
        $s_{LCO\,z'\,20151111 Primary}$ & $0.00290\pm0.00016$\\
        $s_{LCO\,z'\,20151112 Primary}$ & $0.00201\pm0.00097$\\
        $s_{LCO\,z'\,20151119}$ & $0.00135\pm0.00080$\\
        $s_{LCO\,z'\,20151120}$ & $0.00130\pm0.00088$\\
        $s_{Magellan\,z'\,20151120}$ & $0.00116\pm0.00043$\\
        \hline
        \emph{Derived eclipse depths}\\
        $F_p/F_\star$ $z'$-band (\%) & $0.0682 \pm 0.0099 $\\ 
        $F_p/F_\star$ $Ks$-band (\%) & $0.149 \pm 0.014$\\ 
        \hline
    \end{tabular}
\end{table}

\subsection{Independent analysis of each eclipse dataset}
\label{sec:individual-fit}

We also fit each of the eclipse data sets independently to check for self-consistency in our observations. The light curve fitting procedure is similar to that described in Section~\ref{sec:global-fit} and \citet{2015MNRAS.454.3002Z}. The eclipse is modelled with the \emph{JKTEBOP} \citep{1981AJ.....86..102P,2004MNRAS.351.1277S} implementation of the \citet{1972ApJ...174..617N} model. In each case, the free parameters are the surface brightness ratios $S_{p}/S_\star$ and the eccentricity parameter $e\cos w$ that governs the phase of the eclipse. In order to propagate the uncertainties of the planet parameters, the transit geometry parameters are also included in the analysis, including planet-star radius ratio $R_p/R_\star$, normalised orbital distance $(R_p+R_\star)/a$, inclination $i$, orbital period $P$, and transit centre timing $T_0$, but constrained by tight Gaussian priors based on their literature values \citep{2009ApJ...707..167S}. In addition, we allow for simultaneous de-trending any combination of the external parameters $X$, $Y$ target star pixel position, airmass $a$, background flux $b$, target star FWHM $f$, and a linear time trend $t$. 

To choose the right set of de-trending models, we minimise the BIC for the best fit parameters. 
For the 2014-09-25 AAT $Ks$ data set, the eclipse depth is larger if the airmass term is not included in the model (which yields a smaller BIC). We find no large dependence between eclipse depth and de-trending model selection for all other datasets. The derived parameters for each eclipse data set from the independent analysis are presented in Table~\ref{tab:fitparams}.

Tidal decay of the orbital period has been predicted for short period Jovian planets \citep[e.g.][]{2016A&A...588L...6M}. \citet{2014MNRAS.440.1470B} predicted the orbital period of WASP-18b should shorten by $\sim 6$ minutes over a 10 year timescale, whilst recent refinements of the transit ephemeris by \citet{2017ApJ...836L..24W} failed to find any convincing evidence of orbital decay. We make use of our LCOGT primary transits to search for evidence of tidal decay in WASP-18b. We independently fit for the primary transits from LCOGT via the \citet{2002ApJ...580L.171M} model, with a fixed orbital period adopted from \citet{2009ApJ...707..167S}, and derive a transit centroid time of $\mathrm{TDB-BJD} = 2457338.63043 \pm 0.00040$. The transit time is consistent with a linear propagation of the ephemeris from \citet{2009ApJ...707..167S}, and with no appreciable orbital decay. 


\begin{table*}
{\scriptsize
\caption{\label{tab:fitparams} Best fit parameters from independent analyses of each transit}
\begin{tabular}{lrrrrrrrl}
    \hline \hline
    & Period (d) & $T_0$ (BJD-TDB) & $(R_p+R_\star)/a$ & $R_p/R_\star$ & inc ($^\circ$) & $e\cos \omega$ & $S_p/S_\star$ & Detrending  \\
    & & & & & & & &  parameters\\
    \hline
    AAT $Ks$ 20140905 & $0.9414519\,\left(4 \right)$ & $2455084.79293\,\left(9 \right)$ & $0.302 \pm 0.009$ & $0.097 \pm 0.001$  & $84 \pm 2$ & $0.004 \pm 0.005$ & $0.14 \pm 0.03$  & $t$\\
    AAT $Ks$ 20150927 & $0.9414519\,\left(4 \right)$ & $2455084.79293\,\left(9 \right)$ & $0.311 \pm 0.007$ & $0.097 \pm 0.001$ & $86 \pm 1$ & $0.004 \pm 0.003$ & $0.17 \pm 0.02$ & $t,y$\\
    LCOGT  $Z$ 20151102 & $0.9414518\,\left( 4 \right)$ & $2455084.79293\,\left( 9 \right)$ & $0.306 \pm 0.009$ & $0.097 \pm 0.001$ &  $85 \pm 2$ & $-0.004 \pm 0.03$ & $0.07 \pm 0.03$ & $t$\\
    LCOGT  $Z$ 20151104 & $0.9414518\,\left( 5 \right)$ & $2455084.79292\,\left( 9 \right)$ & $0.304 \pm 0.009$ & $0.097 \pm 0.001$ & $85 \pm 2$ & $-0.001 \pm 0.006$ & $0.06 \pm 0.03$ &  $t$\\
    LCOGT  $Z$ 20151119 & $0.9414518\,\left( 4 \right)$ & $2455084.79293\,\left( 9 \right)$ & $0.306 \pm 0.009$ & $0.097 \pm 0.001$ & $85 \pm 2$ & $-0.03 \pm 0.05$ & $0.03 \pm 0.03$ & $t$\\
    LCOGT  $Z$ 20151120 & $0.9414518\,\left( 4 \right)$ & $2455084.79294\,\left( 9 \right)$ & $0.307\pm0.009$ & $0.097\pm0.001$ & $85\pm 2$ & $-0.01_{-0.02}^{+0.05}$ & $0.05 \pm 0.05$ &  $t$\\ 
    Magellan $Z$ 20160917 & $0.9414518\,\left(4 \right)$ & $2455084.79293\,\left( 8 \right)$ & $0.298 \pm 0.008$ & $0.097 \pm 0.001$ & $83 \pm 1$ & $-0.003 \pm 0.002$ & $0.143 \pm 0.005$ &  $t,f$\\    
    \hline
\end{tabular}}
\end{table*}

\section{Planetary Atmosphere Modelling}
\label{sec:results}
 
WASP-18b is separated by only 0.02047\,AU from its F-type host star with $T_{\mathrm eff}=6368$\,K \citep{2012ApJ...757..161T}, which gives a predicted average equilibrium temperature for the planet of just $T_{\mathrm eq}=2410$\,K.  
Dayside observations obtained by \citet{2011ApJ...742...35N} with the IRAC camera on Spitzer, however, are broadly consistent with a blackbody spectrum of $T_{\mathrm rad} = 3200$\,K, which is well above the temperature derived from the expected average dayside insolation. This suggests that majority of the radiation we observe from the planet is received from a region near the sub-stellar point, with a limited (or even absent) energy redistribution \citep{2017A&A...600A..10M}.
\color{black} The lack of significant offset (within 10$^{o}$ error margin) \color{black} between maximum in the thermal phase curve and the time of mid-eclipse in the measurement of phase curve amplitude from the full orbit at 3.6 and 4.5\,$\mu$m \citep{2013MNRAS.428.2645M} also supports the idea of very inefficient redistribution of heat and a low albedo for the planetary atmosphere.

In hot Jupiters, CO and H$_{2}$O molecules are expected to be the primary sources of atmospheric opacity as revealed by prominent absorption features in infrared spectra. The 4.5\,$\mu$m band coincides with strong CO absorption, but the IRAC data in this band for WASP-18b appears to be excessive when compared with the flux measured with other IRAC channels. This was recognised as a possibly revealing the existence of a thermal inversion in the atmosphere of WASP-18b. Thermal inversions had previously been proposed to exist in the atmospheres of the hottest of hot-Jupiters \citep[e.g. HD 209458b, ][]{2007ApJ...668L.171B}. The nature of the absorbers causing such inversions is still debated, with TiO and VO proposed by \citet{2006ApJ...642..495F} and the sulphuric products of photochemistry suggested by \citet{2009ApJ...701L..20Z}. The relatively low level of stellar activity in the WASP-18 host star suggests that any inversion layer formed in the planetary atmosphere will not be easily destroyed by a high intensity stellar UV flux \citep{2010ApJ...720.1569K}.   

However, the observational uncertainties of the Spitzer data permit a reasonably good fit to the spectrum with an atmospheric profile that did not include an inversion, but instead a very gradual increase in temperature with increasing pressure and resulted in shallow absorption features  \citep{2011ApJ...742...35N}. Such a model requires a significantly different chemical composition of the atmosphere. Indeed \citet{2012ApJ...758...36M} showed that atmospheres without requiring inversion and C/O ratios above Solar can yield spectra with very shallow absorption features of oxygen-bearing species like water, as well as strong absorption bands of CO, that fit the spectra of many hot Jupiters. For WASP-18b, the model without inversion predicts rather higher fluxes observed in the near-infrared, which were found to be inconsistent with recently published observations from  HST/WFC3  \citep{2018ApJ...855L..30A,2017ApJ...850L..32S} over the spectral range 1.12-1.65\,$\mu$m. Our observations extend the coverage of near-infrared spectroscopy for WASP-18b from below 1$\mu$m to 2.1$\mu$m.


\subsection{VSTAR atmosphere models}
\label{sec1}
We used the line-by-line radiative-transfer code VSTAR \citep{2012MNRAS.419.1913B} to obtain an independent fit to combined WASP-18b spectral data by testing a wide range of atmospheric pressure and temperature (P-T) profiles, both with, and without, inversions. We created plane-parallel atmosphere models with 44 layers, and derived the atmospheric composition for each layer using the Ionization and Chemical Equilibrium  (ICE) package of VSTAR under the assumption of thermochemical equilibrium and an assumed metallicity and the C/O ratio for the atmosphere. This package accesses a database of 143 compounds in gaseous and condensate phases to derive the abundances of chemical ingredients in terms of mixing ratios for specified molecules that are likely to exist in atmospheres of hot Jupiters. We included the line absorption of 16 molecular and atomic species in the atmosphere of WASP-18b: H$_{2}$O, CO, CH$_{4}$, CO$_{2}$, C$_{2}$H$_{2}$, HCN, TiO, VO, Na, K, Rb, Cs, CaH, CrH, MgH, and FeH. \color{black} The chemical model includes the thermal dissociation of species such as H$_{2}$O and H$_{2}$ in the upper atmosphere, which was shown by \citet{2018A&A...617A.110P} to be an important effect in ultra hot Jupiters. \color{black}

Following a procedure described in \citet{2014MNRAS.445.2746Z} we perform multiple-scattering, radiative-transfer calculations for each layer of the atmosphere on a grid of wave numbers with a specified spectral resolution. A final spectrum is formed by calculating opacity due to molecules and atomic species listed in the comprehensive data base of spectral lines \citep{2012MNRAS.419.1913B}. Additional sources of opacity included in the model are Rayleigh scattering by H, He and H$_{2}$ in the atmosphere, collisionally induced absorption due to H$_{2}$-H$_{2}$ and H$_{2}$-He and the free-free and bound-free absorption from H, H$^{-}$ and H$_{2}^{-}$. The references for the database of absorbers used in our models are listed in Table 2 of \citet{2013ApJ...774..118Z}. The spectrum of the WASP-18 star is derived from the STScI stellar atmosphere models \citep{2004astro.ph..5087C}. 

As mentioned previously the modelling of HST/WFC3 observations led to two different interpretations of atmospheric structure and composition as presented by \citet{2018ApJ...855L..30A} and \citet{2017ApJ...850L..32S}. Although both models require strong thermal inversion in the upper atmosphere,  \citet{2017ApJ...850L..32S} needed a strong reduction in temperature at 1 bar to explain the presumed observation of CO in absorption at 1.6\,$\mu$m and in emission at 4.5\,$\mu$m. Most strikingly their model also requires an unusually high metallicity for C ($C/H=283\times[C/H]_{\odot}$) with an enhanced $C/O$ ratio compared to Solar. It is not clear how such a high discrepancy between the planetary and stellar metallicity \citep{2012ApJ...757..161T} could be achieved in the evolutionary scenario of WASP-18b.  

\begin{figure}
    \centering
    \includegraphics[width=9cm]{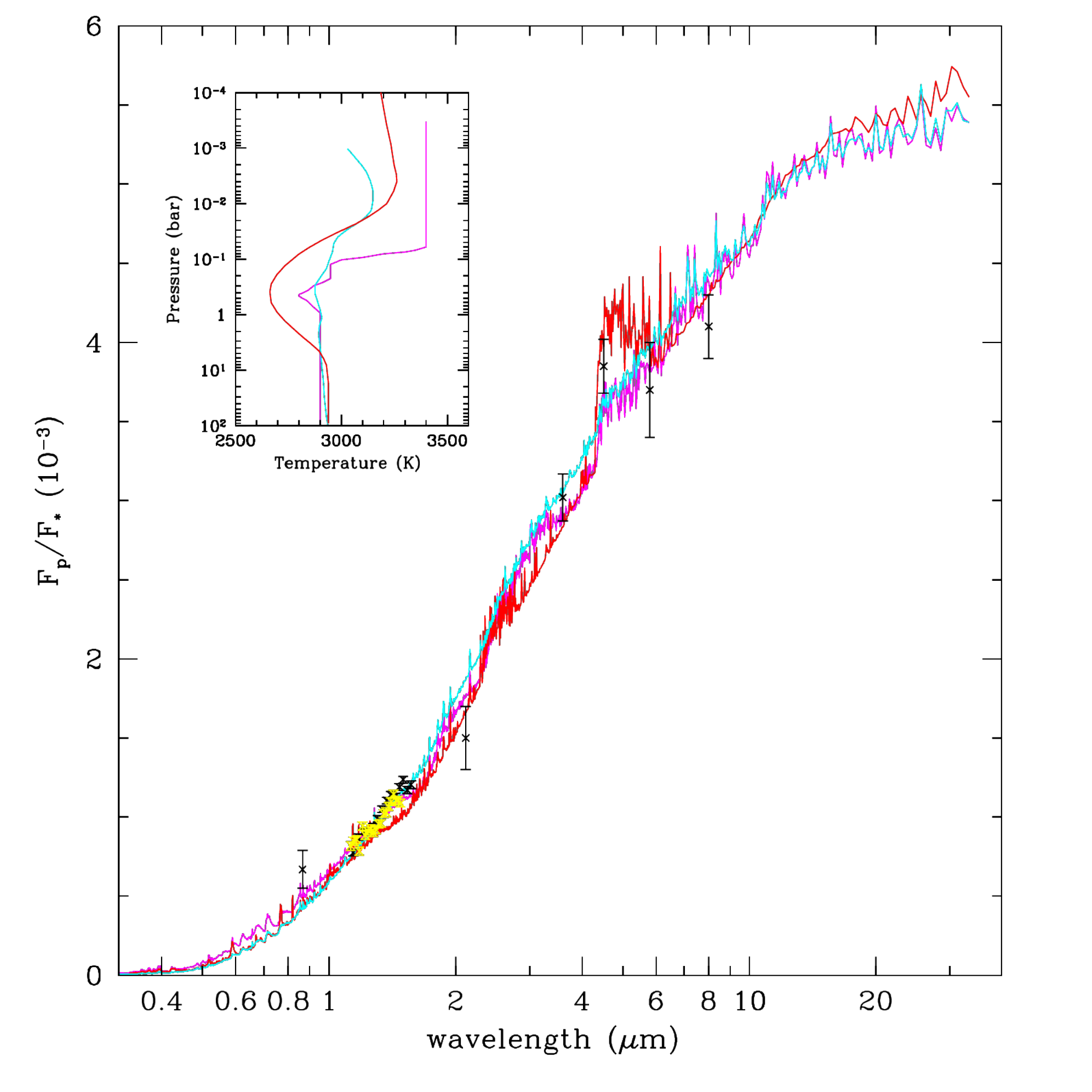}
    \caption{The emission spectra of WASP-18b modelled with the VSTAR code as described in Section~\ref{sec1} (colour-coded). Best-fitting model with C/O=0.54 and Solar metallicity required inversion in top of the atmosphere (in magenta). Models that used P-T profiles from \citet{2018ApJ...855L..30A} (in cyan) and \citet{2017ApJ...850L..32S} (in red). Yellow data points are reproduced from the \citet{2018ApJ...855L..30A} }
    \label{fig:models}
\end{figure}

We tested a variety of simplified pressure-temperature (P-T) profiles, both with and without inversion, and could not fit simultaneously the anomalous Spitzer data point observed at 4.5\,$\mu$m and the $Ks$-band observations. A \chisq-minimisation was used to fit simultaneously both  the previously published data \color{black} from IRAC/Spitzer, WFC3/HST and our new data presented here. \color{black} We found a best-fit model with an inversion in the P-T profile at the level of 0.9\,bar, assuming Solar metallicity and a Solar $C/O$ ratio. This fit is a reasonable match to observations below the $Ks$-band, but is less so for the Spitzer data in the far-infrared that include the anomalously high flux observed at 4.5\,$\mu$m. However our fit does provides a match for \color{black} all four Spitzer data points within their 1-sigma measurement uncertainties. \color{black} We also tested models with $C/O$ ratios between 0.4 and 1.0 while varying atmospheric profiles, and found that the best fitting models matching data in one specific region of the spectrum,  would produce discrepancies in other parts of the spectrum with no significant improvement in \chisq. Clearly more data is needed to resolve this degeneracy in the models. Since we found no compelling reason for a $C/O$ ratio different from  Solar , we consider only such a model (presented in Figure~\ref{fig:models}) in our further discussion.

In Figure~\ref{fig:models}, two other models are shown derived using the P-T profiles of \citet{2018ApJ...855L..30A} and \citet{2017ApJ...850L..32S}. Here we assumed the metallicity and $C/O$ ratios  published by both these groups. The difference in our approach was to derive atmospheric composition using the ICE package for the 16 atmospheric constituents and the same opacity sources as used in our best fitting model. In this way we allow for variation of mixing ratios of atmospheric species as a function of pressure in these two models, which makes a comparison of all these models more meaningful. The resulting spectrum with the P-T profile, high metallicity and C/O ratio from \citet{2017ApJ...850L..32S} underestimates the flux measured within the HST/WFC3 range. While this model provides a better fit at 4.5\,$\mu$m and to the $Ks$-band data from the AAT, our model is statistically more consistent with the results from \citet{2018ApJ...855L..30A} \color{black} including a marginally better agreement with our observations in $z'$-band. This is to be expected since both our models used a more complete set of opacity sources and assume Solar metallicity. \color{black} 

Both papers present slightly different analyses and resulting data from the same HST observations. Perhaps not surprisingly we find that the \citet{2018ApJ...855L..30A} model (which was constructed using just the shorter wavelength range of the HST data) produces a slightly better fit than our model, if we calculate \chisq\ over just this limited short wavelength range.  However our model represents a better fit when the full extent of the HST data \color{black} from \citet{2017ApJ...850L..32S} is considered. \color{black} Although our P-T profiles are similar at pressures below 0.1 bar, we require significantly lower temperatures in the range between 0.1 and 1 bar to deliver a better agreement with our flux measurement in the $Ks$-band. Our best fitting model requires also much higher temperatures at the top of the atmosphere for a closer match to the 4.5\,$\mu$m {\em Spitzer} data point.

\subsection{VSTAR Models with haze opacity}
\label{sec:haze}
  The observed spectra of exoplanets can be strongly affected by both  scattering and absorption due to the presence of condensates in their atmospheres. 
  Formation of hazes and clouds has been suggested to explain transit data for hot Jupiters that show almost featureless spectra with absorption increasing towards blue in the visible, masking the expected absorption from alkali metals and molecular bands. This absorption exceeds that expected from Rayleigh scattering in clear atmospheres \citep[e.g. in HD\,189733b, ][]{2008MNRAS.385..109P,2013MNRAS.432.2917P}. 
  
  \citet{2016Natur.529...59S} analysed transmission spectra of a large sample of hot Jupiters and found them exhibiting a wide range of diverse atmospheres from clear to cloudy. One recognised effect of hazes formed at the top of the atmosphere in transmission spectra is a reduction in the depth of any features due to atomic and molecular absorption. 
  
  On the other hand, secondary eclipse spectra are a combination of both reflected light and planetary thermal emission. The presence of clouds and/or hazes can manifest itself as an increase in the planetary albedo at visible and near-infrared wavelengths \citep{2003ApJ...588.1121S}. Currently measured albedos for all hot Jupiters \citep{2011ApJ...735L..12D,2014PASA...31...43B}  appear to be low (with a notable exception of Kepler-7\,b). Theoretical considerations also suggest that a high stellar irradiation could prevent a planet from forming clouds similar to these which cause high albedos of the Solar system planets. 
  Only the most refractory species with the highest condensation temperatures -- e.g. corundum (Al$_{2}$O$_{3}$), iron (Fe) and enstitate (MgSiO$_{3}$) along with few other molecules -- could form clouds in the upper layers of such hot atmospheres. Such condensates have also been  suggested as responsible for dust formation in the atmospheres of L-type brown dwarfs \citet{2014ApJ...789L..14M}.

  Particles that form clouds or hazes absorb in different parts of the spectrum, but most typically in the mid-infrared where the molecular vibrational modes are present. These absorption features depend on the optical properties, size, distribution, shape and opacities of the cloud or haze. 
  
\begin{figure*}
    \centering
    \includegraphics[width=19cm]{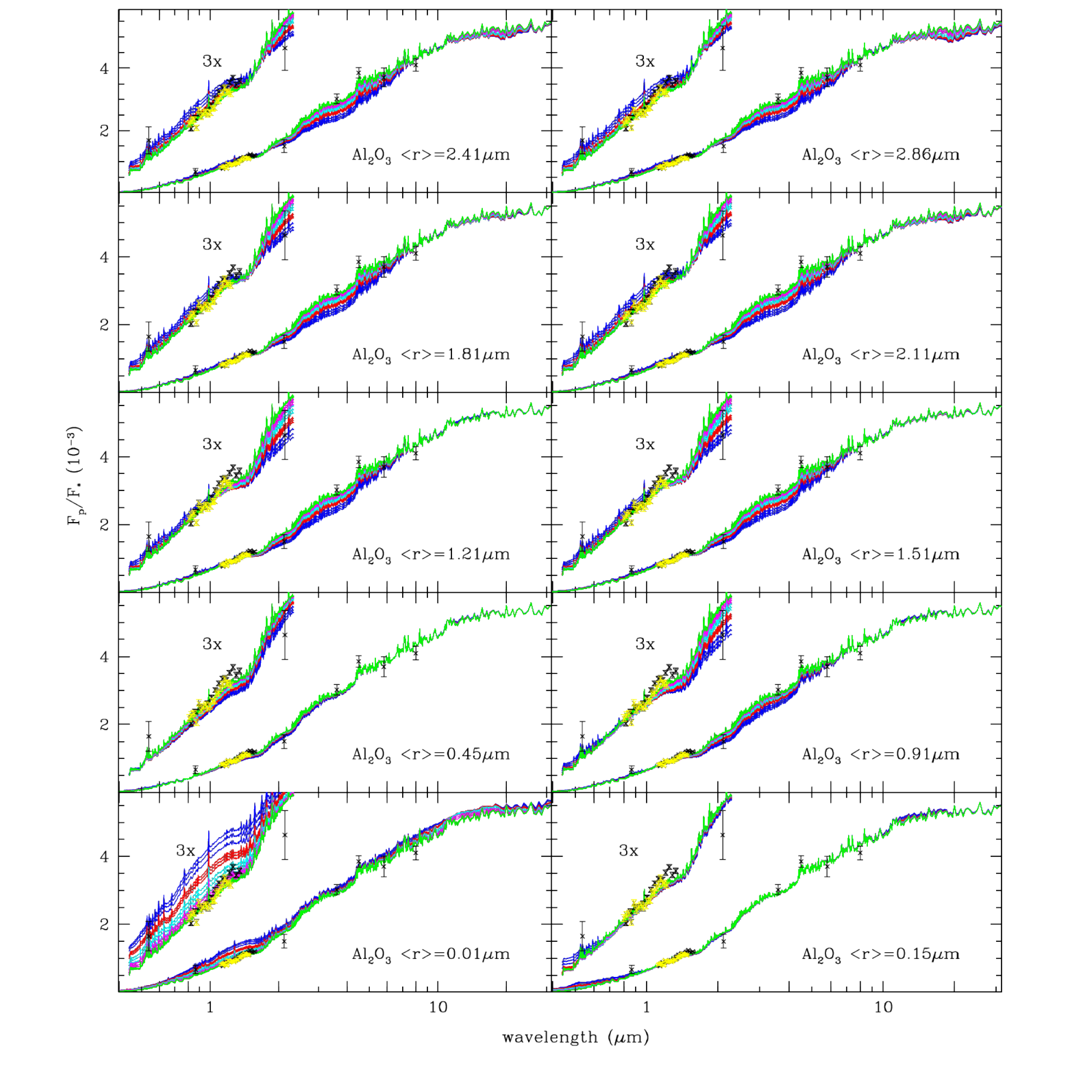}
    \caption{The emission spectra of WASP-18b modelled with the VSTAR code as described in Section~\ref{sec1} with an addition of the haze  with a complex refractive index of Al$_{2}$O$_{3}$ molecules at the top of the atmosphere. Models in each panel are derived for different characteristic sizes of particles $<r>$  (Section~\ref{sec:haze}). There are 15 spectra in every panel each plotted with optical depth between  $\tau$=0.001 (green) to $\tau$=1 (blue). The colour scheme is chosen such that the blue, red, cyan, magenta and green are ordered according to decreasing $\tau$. In each panel we also provided an expanded version of the spectra in the wavelength range between 0.8 and 2.2\,$\mu$m, scaled by a factor of 3.}
    \label{fig:models1}
\end{figure*}

\begin{figure*}
    \centering
    \includegraphics[width=19cm]{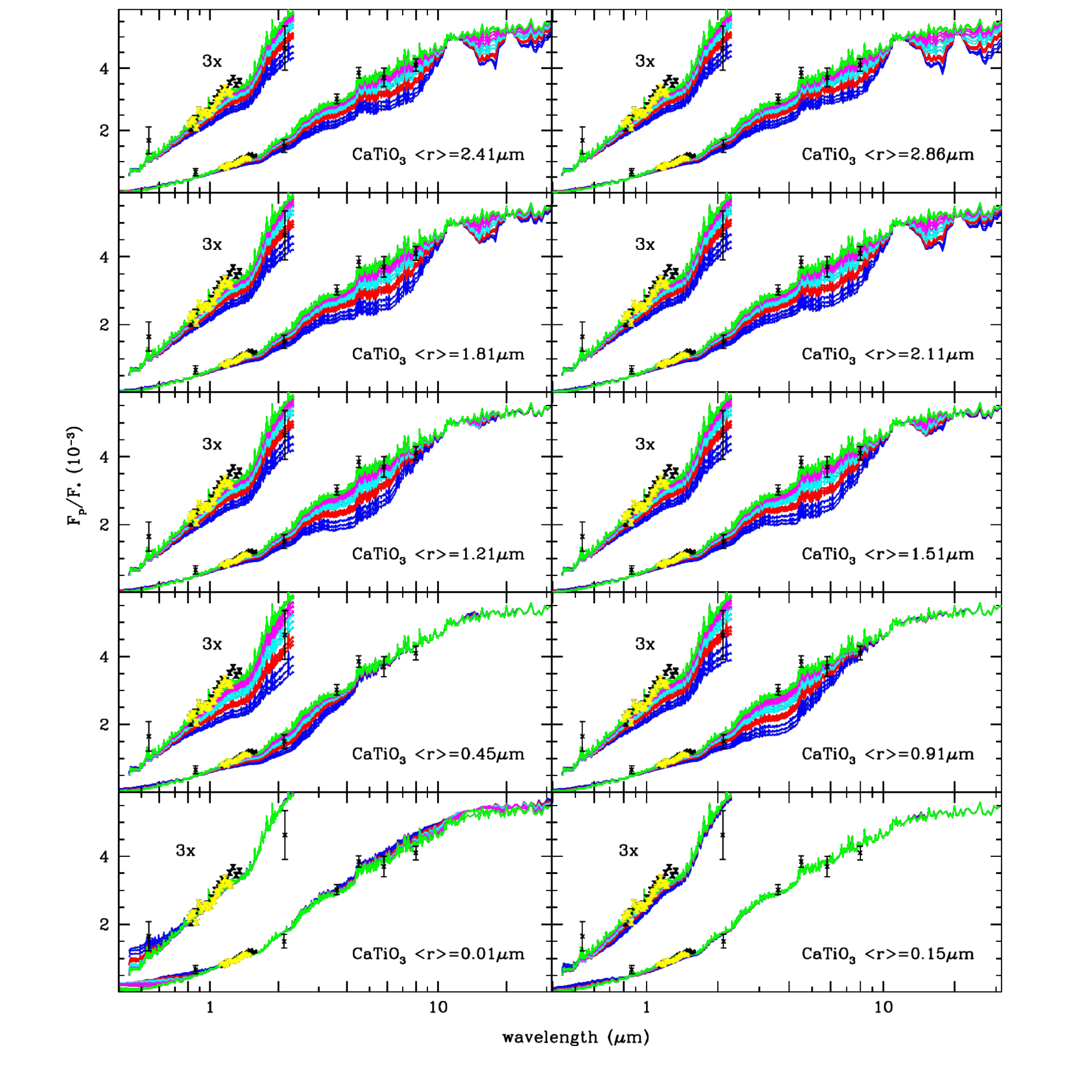}
         \caption{The same as figure~\ref{fig:models1} but particles that form haze have a complex refractive index of perovskite molecules, CaTiO$_{3}$.} 
    \label{fig:models4}
\end{figure*}

\begin{figure*}
    \centering
    \includegraphics[width=19cm]{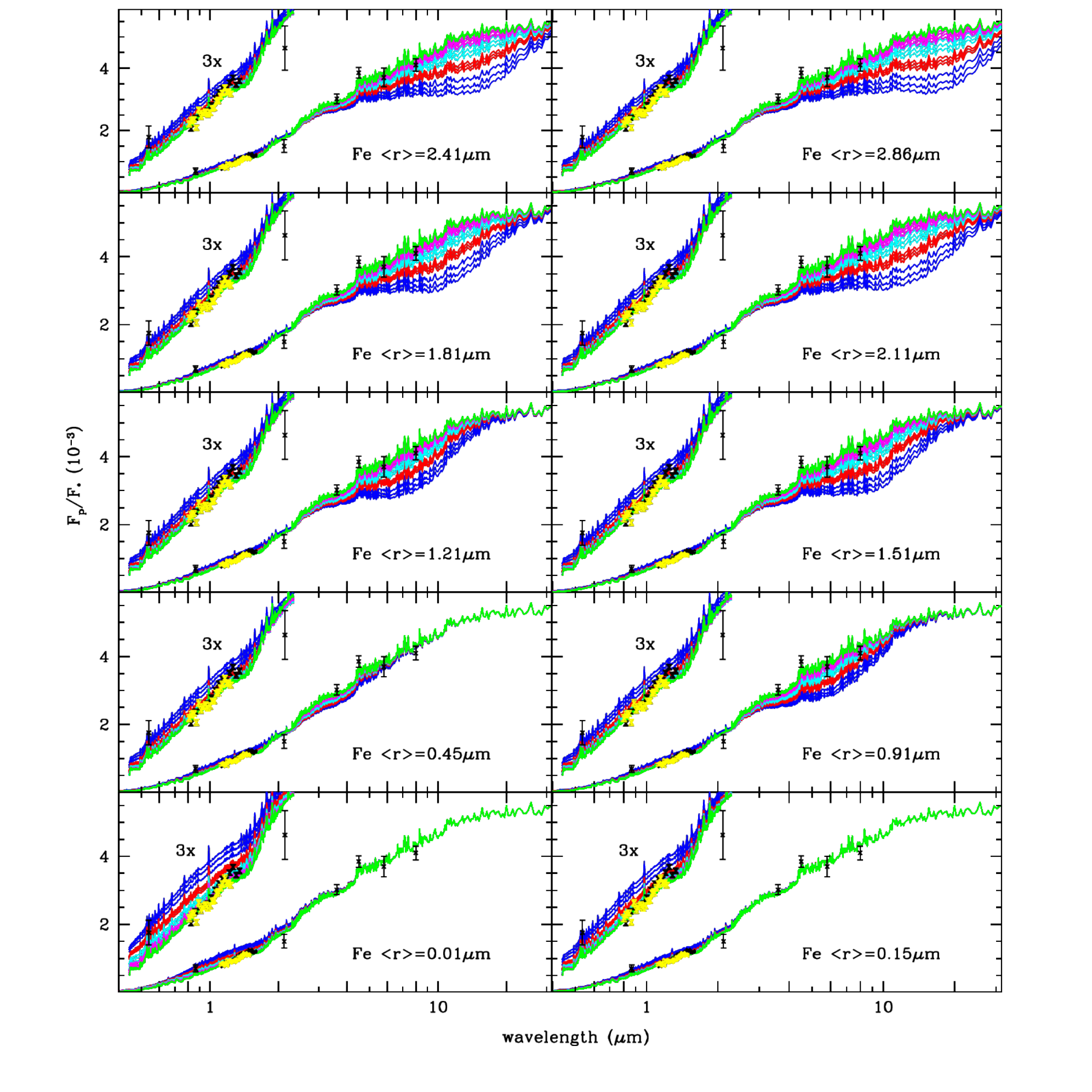}
    \caption{The same as figure~\ref{fig:models1} but particles that form haze have a complex refractive index of solid Fe.} 
    \label{fig:models2}
\end{figure*}

\begin{figure*}
    \centering
    \includegraphics[width=19cm]{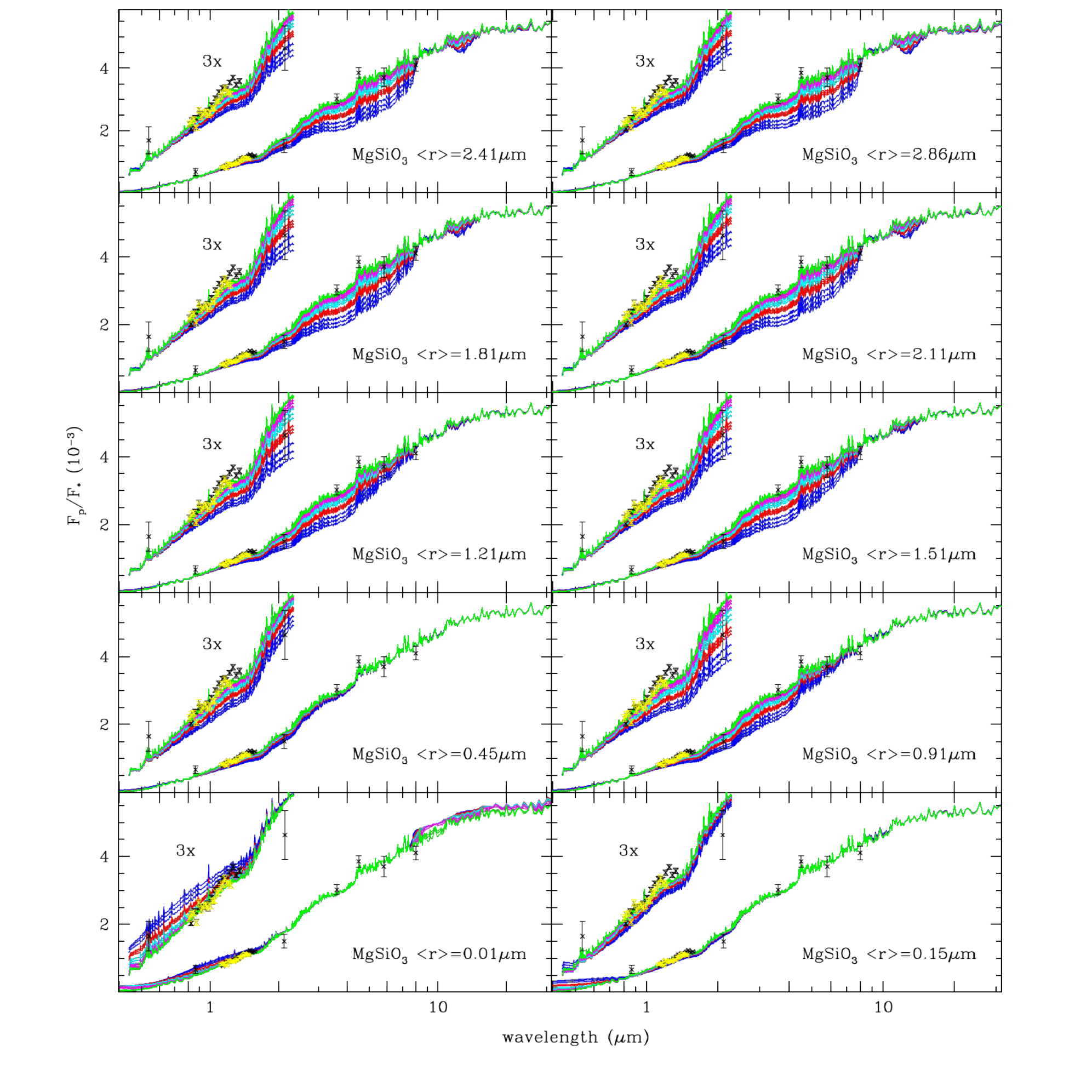}
     \caption{The same as figure~\ref{fig:models1} but particles that form haze have a complex refractive index of enstatite molecules, MgSiO$_{3}$.} 
    \label{fig:models3}
\end{figure*}

  
  We calculated these reflection and absorption effects in the emission spectra for four selected types of condensate in the atmosphere of WASP-18b (Figures~\ref{fig:models1},~\ref{fig:models2},~\ref{fig:models3} and ~\ref{fig:models4}). Currently there are no known species that condense at the temperatures derived for the substellar point of this planet's atmosphere, especially if the presence of thermal inversion is required at pressures lower than 0.1 bar. However significantly lower temperatures are predicted for the night side of WASP-18b in the time dependent radiative transfer solutions of \citet{2013Icar..226.1719I}. They derive thermal profiles, which show up to a 50\% difference in temperature at 0.1 bar between the night- and day-sides. It is not inconceivable that condensation occurs in cooler, night-side temperatures and that clouds are distributed via atmospheric circulation beyond the terminator \citep{2011ApJ...738...71S,2015ApJ...801...95S}, where they gradually evaporate. Any hazes formed via photochemical processes could potentially persist in hotter conditions of the dayside. All this could lead to dynamic and possibly highly variable conditions in the atmosphere of an (at least partially) cloud-covered planet. \citet{2016ApJ...828...22P} estimates the effective cloud coverage at the dayside for the most refractory condensates suggested for hot Jupiters in the temperature range between 1000 to 2200\,K. We considered the top four species discussed in that paper, for which the effective cloud coverage at the dayside was above 0.1 at the highest temperatures modelled there \citep[see Figure 13 in][]{2016ApJ...828...22P}.
  
  In the models presented here we placed a hypothetical haze at a pressure of 0.1 mbar, and varied its optical depth at 1.0\,$\mu$m, and the mean size of its particles using a power law distribution with an effective variance of $v_{eff}=0.02$. The VSTAR code calculations of the emergent fluxes are calculated using Mie theory applied to spherically symmetric particles \citep{2002sael.book.....M}. To calculate models for different cloud molecules, we used the complex refractive indices provided in \citet{2018MNRAS.475...94K} and \citet{2015A&A...573A.122W}. In the panels of Figures~\ref{fig:models1},~\ref{fig:models4},~\ref{fig:models2} and ~\ref{fig:models3} the emission spectra for ten different sizes of cloud particles between $<r>$ = 0.01 and 2.86\,$\mu$m are presented. Each panel shows 15 colour-coded spectra calculated with different optical depths of the cloud between $\tau$=0.001 (in green) to 1 (blue).\footnote{The colour scheme is applied as follows. We plot 3 spectra in each colour: green, magenta, cyan, red and blue with progressively higher optical depths. So the green spectra correspond to low optical depths, while magenta, cyan, red - to increasingly higher optical depths, and blue spectra represent the highest values.} 
  By examining all four plots it is immediately apparent that observations in the near-infrared (like those taken with HST) are relatively insensitive to the presence of any clouds   within investigated range of parameters for optical depth and particle size. 
    
  Figure~\ref{fig:models1} shows impact on  emission spectra of the presence of  Al$_{2}$O$_{3}$ clouds. As expected, modelled spectra show increased reflectivity that extends well into the near-infrared region for the smallest particles. Absorption features extending across the mid-infrared, and increasing with opacity of the cloud, seems to be the most significant impact of Al$_{2}$O$_{3}$ clouds on these emission spectra. New observations, at higher precision than current {\em Spitzer data}, in this region could potentially discern between optically thin and optically thick Al$_{2}$O$_{3}$ clouds. The currently available data fit best to models of the former case. It is worth noting that the presence of optically thick cloud could be helpful in explaining the \color{black} lower observed flux in the $Ks$ band and increased emission in $z'$ band compared with that predicted by clear atmosphere models. \color{black} 
  We also included models for spectra affected by clouds of perovskie (CaTiO$_{3}$) which show a characteristic absorption structure in the far-infrared region for large particles and high opacities, while increasing reflectivity of the planet in the visible spectrum for small particles (Figure~\ref{fig:models4}). 
  Iron condensates affect the spectra in a manner similar to Al$_{2}$O$_{3}$ clouds for the smallest particles (Figure~\ref{fig:models2}). While the increased optical depth causes much deeper absorption in the mid- and far-infrared region, and increased reflectivity in $z'$ band region, it   leaves $Ks$ band largely unaffected.  
 Modelled spectra with clouds composed of enstatite particles with mean sizes larger than 0.5\,$\mu$m (Figure~\ref{fig:models3}) show broad absorption features in the same regions of the near- and mid-infrared as seen for  Al$_{2}$O$_{3}$ clouds.
 
 \color{black} We find the emission in $z'$-band to be sensitive to the increased opacity of hazes made from the particles smaller than 0.15\,$\mu$m. The $z'$-band seems to be of limited value as a diagnostic of cloudiness in the planet as both increased opacity and increased particle size affects that region only slightly in comparison to the near- and mid-infrared spectra.  \color{black}
  
 \section{Discussion} 
 \label{sec:discussion}
 Currently available data for the hot Jupiter  WASP-18b, are not sufficient to derive a unique model for its atmosphere and to resolve degeneracies that arise when we attempt to model too many unknown parameters, such as the detailed thermal profile, chemical composition and the structure of condensates.
 Our cloudy models are derived for a simple, one layer cloud using a single, specific temperature profile of the atmosphere with Solar metallicity and C/O ratio, without addressing any non-equilibrium processes that could have effects on the atmospheric composition and chemistry (for example on the abundance of CO and CH$_{4}$). We have not discussed formation processes for such clouds, which can be reviewed in \citet{2013cctp.book..367M}. We have also neglected the feedback of cloud formation on the P-T profile. 
 
 We present plausible models for the WASP-18b emission spectra with the opacities from clouds and hazes made from some of the most refractory species. These are given as the examples of effects that can make interpretation of data challenging. 
We found that a model with inversion in the upper layers of the atmosphere fits best all the data currently available. However the fact that these multi-wavelength observations were not obtained simultaneously makes any such models flawed, if the variability due to changes in cloud opacity or cloud cover are significant. 
It is still debated if any condensates can survive in the atmosphere of objects as hot as WASP-18b. One potentially viable location for the formation of such clouds is the night side of the planet, where temperatures are expected to drop significantly enough to allow the condensation of the most refractory compounds (such as corundum and solid iron). Any condensates would have to be transported to the dayside of the planet by equatorial super-rotating winds developed in tidally locked hot Jupiters   \citep{2011ApJ...738...71S,2015ApJ...801...95S}. 

The observation of planetary albedos can provide clues as to the presence of any haze composed of  small particles  \citep[as has been implied for HD\,189733b ][]{2013MNRAS.432.2917P}. In addition to phase curve measurements from \citet{2013MNRAS.428.2645M}, our polarimetry observations with the High Precision Polarimeter (HIPPI) will provide independent limits on the planetary albedo of WASP-18b (Bott et al. submitted to AJ).    
As our models show, the most significant effect of clouds on planetary emission spectra can be seen at mid- and far-infrared wavelengths. This is a region which JWST will make accessible. Two JWST instruments, NIRSPEC (0.6-5$\mu$m) and MIRI (5-28$\mu$m), will be able to provide suitable spectra with adjustable resolution. \citet{2017A&A...600A..10M} estimate that just four repeated observations of WASP-18b with NIRSPEC will be able to distinguish between very similar spectra that require thermal inversion, as opposed to the elevated $C/O$ ratio in their model without clouds. Our examples show that the effects of clouds will make the analysis of JWST data more complex, but that the simultaneous coverage of mid- and far-infrared wavelengths will provide necessary information to understand any potential variability of the emission spectrum due to clouds. Partially cloudy models have already been developed \citep{2016ApJ...828...22P} and the combination of eclipse data in the visible, together with the improved spectral coverage from JWST, will be crucial to understanding the composition and the dynamics of atmospheres in the most irradiated planets.

\section{Conclusions}
\label{sec:discussion}
We present new observations of WASP-18b eclipses and transits with ground-based telescopes (the AAT, Magellan and LCOGT) which complement previous observations from space telescopes (HST and Spitzer). Previously published models show that it has been a challenge to understand the observed data with a consistent composition for the WASP-18b atmosphere.

\color{black} Our observations over a broadened near-infrared baseline seem to overestimate emission at $z'$-band compared with clear atmosphere models derived from the HST data, and also to underestimate $Ks$-band emission expected in the model of \citet{2018ApJ...855L..30A}. 
On the other hand, we also could not reproduce an exact match for the HST data with the parameters published by \citet{2017ApJ...850L..32S}, which is less problematic, because their model did not include important contributors to atmospheric opacity, such as H$^{-}$ and the effects of water dissociation. \color{black} We applied simple cloud models to our best fit model to show the possible effects of varied cloud particle sizes and opacities in different parts of the spectrum. It is plausible that the \color{black} lower flux in $Ks$-band could be a result of the absorbing, high altitude haze \color{black} composed of particles with relatively large sizes. Such particles could be uplifted in the hot atmosphere, which is expected to show vigorous mixing. \color{black} Such clouds could also explain increased reflectivity in the $z'$-band as seen in Figure~\ref{fig:models1}. 

\color{black} Ground-based broadband observations can generally act as precursors to more detailed characterisation by HST and Spitzer, and understanding how they match with the eventual space-based observations is key to placing these precursors into context. For example the deep $z'$-band eclipses of KELT-1b \citep{2012ApJ...761..123S} raised questions on the heat redistribution on the brown dwarf, eventually leading to the a phase curve by Spitzer/IRAC \citep{2014ApJ...783..112B,2018arXiv180809575B}. 
 \color{black}
 Given the current questions regarding the longevity of HST and Spitzer, it is increasingly important to use the benchmark systems like WASP-18b, to guide the interpretation of future ground-based broadband secondary eclipse observations.\color{black}

\section*{Acknowledgements}
\label{sec:acknowledgements}
We thank the support of the AAO staff who helped with establishing the observing strategy employed in this work. CGT gratefully acknowledges the support of ARC Discovery Outstanding Researcher Award DP130102695. This paper includes data gathered with the 6.5 meter Magellan Telescopes located at Las Campanas Observatory (LCO), Chile. D.D. acknowledges support provided by NASA through Hubble Fellowship grant HSTHF2-51372.001-A awarded by the Space Telescope Science Institute, which is operated by the Association of Universities for Research in Astronomy, Inc., for NASA, under contract NAS5-26555.

\newcommand{\newblock}{}

\bibliographystyle{mn2e}
\bibliography{mybib.bib}


\label{lastpage}

\end{document}